\documentclass[11pt]{article}

\usepackage{amsmath, amssymb, enumerate, ./fullpage}
\usepackage{graphicx}
\usepackage{mdframed, xcolor}

\usepackage{hyperref}

\newtheorem{theorem}{Theorem}
\newtheorem{lemma}{Lemma}

\def\beginProof{\par{\it Proof}. \ignorespaces}
\newcommand{\closeProof}{\hfill $\blacksquare$}

\newcommand{\beginE}{\begin{equation}}
\newcommand{\beginEL}[1]{\begin{equation}\label{#1}}
\newcommand{\closeE}{\end{equation}}

\newcommand{\beginPM}{\begin{pmatrix}}
\newcommand{\closePM}{\end{pmatrix}}

\newcommand{\realR}{{\mathbb{R}}}

\newcommand{\grad}{\nabla}

\newcommand{\inner}[2]{\big\langle #1, \, #2 \big\rangle}

\newcommand{\dt}{\Delta t}

\newcommand{\myand}{\qquad \mbox{and} \quad}

\newcommand{\mywith}{\qquad \mbox{with} \quad}

\newcommand{\E}{\mathrm{E}}
\newcommand{\EXP}[2]{\E_{#1}\left[#2\right]}	

\newcommand{\Var}{\mathrm{Var}}

\newcommand{\bx}{{\boldsymbol x}}

\newcommand{\bv}{{\boldsymbol v}}

\newcommand{\bh}{{\boldsymbol h}}
\newcommand{\bbeta}{{\boldsymbol \beta}}

\newcommand{\ba}{{\boldsymbol a}}
\newcommand{\bb}{{\boldsymbol b}}	
\newcommand{\bz}{{\boldsymbol z}}	
\newcommand{\bu}{{\boldsymbol u}}	
\newcommand{\bW}{{\boldsymbol W}}	

\newcommand{\Prob}{\mathrm{Prob}}

\newcommand{\eps}{\varepsilon}

\begin{document}

\setlength{\baselineskip}{12.8pt}

\title{ A Stochastic LQR Model for Child Order Placement \\ in Algorithmic Trading 	
        \thanks{Dr. Jackie Shen has been the Global Project Leader for Algo Trading Model Risk at Goldman Sachs, New York, USA. Previously, he had also been a quantitative strategist at the equities Algorithmic Trading desks at both J.P. Morgan and Barclays, New York, USA. Email: jhshen@alum.mit.edu; URL: alum.mit.edu/www/jhshen }           }

\author{ Jackie Jianhong  Shen 		\\ \\
		 Financial Services			\\
         New York City, USA 	\\
        \\
        \\
        March 25, 2020
        }

\date{}

\maketitle

\begin{abstract}
Modern Algorithmic Trading (``Algo") allows institutional investors and traders to liquidate or establish big security positions in a fully automated or {\em low-touch} manner. Most existing academic or industrial Algos focus on how to ``slice" a big parent order into smaller child orders over a given time horizon. Few models rigorously tackle the actual placement of these child orders. Instead, placement is mostly done with a combination of empirical signals and heuristic decision processes. A self-contained, realistic, and fully functional Child Order Placement (COP) model may never exist due to all the inherent complexities, e.g., fragmentation due to multiple venues, dynamics of limit order books, lit vs. dark liquidity, different trading sessions and rules. In this paper, we propose a reductionism COP model that focuses exclusively on the interplay between placing passive limit orders and sniping using aggressive takeout orders. The dynamic programming model assumes the form of a stochastic linear-quadratic regulator (LQR) and allows closed-form solutions under the backward Bellman equations. Explored in detail are model assumptions and general settings, the choice of state and control variables and the cost functions, and the derivation of the closed-form solutions.

\vskip 16pt

	
\vskip 24pt
\begin{mdframed}[backgroundcolor=orange!20] 
\noindent \textbf{Attention:} \; The current work is designed to be exclusively published in the Social Sciences Research Network (SSRN) or arXiv preprint server. Its commercial or open-journal publication is prohibited without the prior consent of the author. The current free style accommodates informative cover pages and colored text boxes that help enhance the reading experience. 
\end{mdframed}

\end{abstract}	
	

\newpage
\tableofcontents

\newpage 
The following terms are frequently used throughout the paper. 
\begin{table}[!htbp]
\centering
\label{tbl:list_terms}
\colorbox{blue!8}{
	\begin{tabular}{|l|l|}\hline
	 \textbf{term}  & \textbf{meaning} 			\\ \hline
	 Algo 	& an automated trading process based on algorithms \\ \hline
	 Profile & an intraday time series derived from history, e.g., for volumes \\ \hline
	 IS 	& implementation shortfall - a popular optimization-based Algo  \\ \hline
	 TWAP 	& time-weighted average price - a standard profile-based Algo \\ \hline
	 VWAP 	& volume-weighted average price - a standard profile-based Algo \\ \hline
	 COP 	& child order placement  			\\ \hline
	 DP 	& dynamic programming  				\\ \hline
	 LQR 	& linear quadratic regulator with linear transitions \& quadratic costs 		\\ \hline
	 (N)BBO 	& (national) best bid and offer of a venue or market system  \\ \hline
	 LOB	& limit order book, with buy/sell orders displayed \\ \hline
	 SOR	& smart order router/routing		\\ \hline
	 FX 	& foreign exchanges 				\\ \hline
	 ADV 	& average daily volume 				\\ \hline	 
	 Passive Touch & best bid for buy and best offer for sell \\ \hline
	 Aggressive Touch & best offer for buy and best bid for sell \\ \hline
	 Near Touch & same as Passive Touch 		\\ \hline
	 Far Touch  & same as Aggressive Touch 		\\ \hline
	 Takeout or Sniping & aggressive market order at the far touch \\ \hline
	 TMV 	& true market value 				\\ \hline
	 Positive Matrix  & a symmetric real matrix with positive eigenvalues \\ \hline
	\end{tabular}
}
\end{table}

The following symbols have been consistently used in the paper. 
\begin{table}[!htbp]
\centering
\label{tbl:list_symbols}
\colorbox{blue!8}{
	\begin{tabular}{|l|l|}\hline
	 \textbf{symbol}  & \textbf{meaning} 						\\ \hline
	 $t_n$ 		& a discrete action time in dynamic programming 		\\ \hline
	 $X_n^\pm$ 	& the right/left limit of a quantity $X$ at $t_n$, via $t_n\pm\eps$ \\ \hline
	 $HS$  		& half spread between the BBO 	\\ \hline
	 $Mid$ 		& mid price between the BBO 		\\ \hline
	 $N(\lambda)$ 	& Poisson distribution with rate $\lambda$  \\ \hline
	 $q_n$ 		& outstanding positions right before $t_n$; a state variable  	\\ \hline
	 $\lambda_n$& Poisson hitting rate right before $t_n$; a state variable 	\\ \hline
	 $u_n$ 		& aggressive market orders at $t_n$; the control variable 		\\ \hline
	 $\eta$ 	& market impact of aggressive orders on passive fills			\\ \hline
	 $\gamma_n$	& delay cost penalty at $t_n$; also expressing risk aversion \\ \hline
	 $\bx_n$	& state variable right before $t_n$; $\bx_n=(q_n, \lambda_n)$		\\ \hline
	 $V_n(\bx_n)$ & the value function at $t_n$ in dynamic programming \\ \hline
	\end{tabular}
}
\end{table}


\newpage
\section{Introduction to Child Order Placement (COP)}
\label{sec:1-intro}
Small retail orders can be filled by simple vanilla market or limit orders. There is no need for  automated algorithmic trading (or ``Algos") involving sophisticated strategies. Algos are primarily designed to execute sizable institutional orders from portfolio managers or traders in various funds or broker-dealers. In the current work, the term ``Algos" is restricted to the automated execution service provided to either external or internal clients by the buy side, sell side, or specialized execution agents. 

A typical single-name Algo is stratified to at least two distinguishable layers, which we shall refer to as ``macro" and ``micro" layers or (sub-)Algos. 
\begin{enumerate}[(A)]
\item At the macro layer, a big parent order is sliced into smaller child orders over a series of time buckets, e.g., 5-minute intervals, which usually depend on the liquidity profiles of a security.

\item The micro layer handles the execution of the resulted child orders. The actual implementation and architecture could vary significantly among broker-dealers and execution agents, and constitute into the very proprietary core of all Algos. 
\end{enumerate}

Most well-known Algos in the industry are named after the macro layers, e.g., TWAP, VWAP, or Implementation Shortfall (IS). These macro Algos are either configured based on historical benchmarks or optimized under proper utility objectives. The optimization techniques involve either static or dynamic frameworks, as in these sample works~\cite{algo_almgren12,algo_bertsimas_Lo98,algo_bouchard_dang_lehalle2011,algo_hora2006_lqr,
algo_huberman_stanzl01,algo_shen15_pretrade,algo_shen17_lqrIS,algo_shen15_mvmvp}. Overall, the macro Algos are built upon the macro behaviors of the targeted securities, including for instance, the historical profiles of volumes, volatilities, spreads, etc, as in our earlier works in~\cite{algo_shen15_pretrade,algo_shen17_lqrIS,algo_shen15_mvmvp}. They do not act upon the real-time micro structure signals such as the dynamics of limit order books (LOB).

The complexities of these Algos mainly reside within the micro layers. The actual real-time placement of orders on various venues is implemented at this layer, and different Algo providers may take very different approaches. For example, two  offerings of the same VWAP Algo could differ significantly in terms of architecture and logic. 

In general, a micro Algo must handle actions like the following:
\begin{enumerate}[(a)]

\item a dynamic decision flows for the placement actions and monitoring of their status,

\item allocation among different order types offered by all accessible venues, and

\item real-time routing to all accessible liquidity venues, including both lit and dark venues. 
\end{enumerate}
In particular, the last component often assumes its own identity in most broker-dealer Algo offerings, and is called the SOR - Smart Order Router. SORs are vital for some liquid asset classes with highly fragmented markets, e.g., the common stocks in the USA. 

In terms of modeling techniques, macro Algos are either configured or scheduled using benchmark profiles such as TWAP or VWAP, or optimized using proper utility functions (e.g., the mean-variance framework of Almgren-Chriss~\cite{algo_almgren_chriss00}; also see Shen et al.~\cite{algo_shen15_pretrade,algo_shen17_lqrIS,algo_shen15_mvmvp}). Modeling of micro Algos is much more challenging due to the aforementioned multiple tasks. The main complexities inherent to market microstructures include  venues, sessions, order types, and their optimal real-time management. SOR is part of this grand effort and probably the most well known or actively marketed by Algo providers. But the SOR alone is only the last segment of the placement stream, and is actually irrelevant for single-venue securities such as commodities futures or some FX products.

In this paper, we attempt to develop a COP model for a single-venue security. Hence SOR is out of the scope. Instead, the primary focus is on how to dynamically place and manage aggressive market  orders and passive limit orders. The two order types compete in the following manner.
\begin{enumerate}[(i)]
\item Aggressive market orders get filled fast and hence help accomplish order completion, but at the cost of paying a half spread (with respect to the market mid price) and information leakage that could harm trailing orders. 

\item Passive limit orders gain a half spread if filled, but are subject to fill uncertainty and an extra ``chasing" cost if the market drifts away. In general, they jeopardize order completion. 
\end{enumerate}
A COP Algo in this context must optimally manage the interplay between the two order types, in order to achieve the two primary objectives:
\begin{enumerate}[(a)]
\item lower total cost for filled orders, and

\item a higher completion rate.
\end{enumerate}

The current work rolls out as follows. 
After making some reductionism assumptions, we first build the COP model based on stochastic dynamic programming (SDP). The solution to the resultant stochastic linear-quadratic regulator (LQR) problem is then worked out via the associated Bellman equations. Theoretical assumptions or practical implications are always discussed in detail along the way.  The main result is summarized into Theorem~1 in Section 4. 

We believe this is the first self-contained and mathematically rigorous COP model that has a closed-form solution.

\section{Basic Model Settings and Scopes}
\label{sec:2-setting}

\subsection{From Macro to Micro}
\label{subsec:2.1}

Throughout the rest, in order to gain a more tangible sense of all the model settings, we assume a concrete working parent order with the following attributes.
\begin{enumerate}[(i)]
\item It is a ``buy" order for a common stock, say.
\item The total quantity is $Q=100,000$ shares.
\item The average daily volume (ADV) is $2,000,000$ shares, based on a monthly rolling window.
\item The client prefers the order to be completed during the horizon from 12:00 pm to 3:00 pm EST in the US equity market (which is however consolidated into a single-venue market to avoid SOR).
\end{enumerate}
Of course none of these example order details actually puts restrictions on the proposed COP model. For example, the model is applicable to other liquid asset classes such as futures and rates. 

At the macro level, Algos such as TWAP, VWAP and IS ``slice" the parent order into smaller child orders over a series of time buckets. For illustration, assume that such an Algo works with  5-minute time buckets. Then the target 3-hour execution horizon requested by the client is split into 36 time buckets. Further assume that this macro Algo decides to allocate $2,500$ shares or 25 lots to the specific time bucket [1:00pm, 1:05pm]. In practice these time knots can all be randomized for anti-gaming.

If the time buckets of the macro Algo are relatively big, e.g., 5 minutes, a micro-layer scheduler can be further designed to schedule the allocated $2,500$ shares, say, over finer micro bins. At this layer, sophisticated optimization may be spared in order to save time. In general, an equal partition, i.e., allocation based on TWAP, can be applied. Take the above working example for instance. The 25 lots allocated for the macro time bucket [1:00pm, 1:05pm] can be further split to 5 lots over each micro bins of 1 minute, i.e., [1:00pm, 1:01pm], [1:01pm, 1:02pm], etc. Again randomization of time knots should also be applied. 

The introduction of finer micro time bins is necessary for most scenarios. This is because the macro time buckets must last long enough so that bucket signals are statistically meaningful. Take the VWAP or IS Algo for example. These macro Algos all depend on the temporal profiles of security volumes, volatilities, or spreads (e.g., Shen et al.~\cite{algo_shen15_pretrade,algo_shen17_lqrIS,algo_shen15_mvmvp}). When the macro buckets are too brief, the profile values based on historical averaging or filtering would be too noisy and result in unreliable Algo scheduling at the macro layer. In general, bucket size should be optimized or adapted to the liquidity profiles of a given security. 

The current work takes a reductionism approach and attempts to develop a self-contained rigorous COP model at the micro bin level. That is, the model handles the actual execution of individual child (or grandchild) orders over micro bins, e.g., 5 lots over [1:00pm, 1:01pm] for the above running example. It is a fully automated model based on the framework of stochastic linear-quadratic regulators (LQR) and allows closed-form solutions. 

\subsection{Reductionism and Value of the Current Work}
\label{subsec:2.2-reductionism}

In our previous two works on macro Algos:
\begin{itemize}
\item Shen~\cite{algo_shen15_pretrade} on a generic pre-trade macro Algo based on static quadratic programming in Hilbert spaces, and
\item Shen~\cite{algo_shen17_lqrIS} on a real-time adaptive macro Algo based on dynamic programming that integrates the VWAP and IS Algos,
\end{itemize}
the proposed Algo models can actually be implemented in execution houses after proper model calibration is performed using their proprietary trading data. 

Hence it is important to point out at the outset that the current micro Algo is more a theoretical model in the spirit of reductionism. As explained earlier, it is almost impossible to have a single self-contained model to comprehensively handle the entire micro-layer execution. For instance, it is nontrivial to handle unexpected intraday trading halts or participation in various auctions. The main reductionism of the proposed model involves the following aspects.
\begin{enumerate}[(a)]
\item It is restricted to single-venue executions and does not involve SOR modeling.

\item It deals with neither lit vs. dark venues nor complex order types (e.g., icebergs or mid-pegging).

\item It only handles the two most basic order types - limit and market orders. 
\end{enumerate}
Notice that in professional trading one almost never sends a ``naked" market order. Hence by market orders we mean more precisely {\em marketable} limit orders whose limits cross the far touch. 

Despite the reductionism, the value of the current work can be summarized as follows.
\begin{itemize}
\item To the academic community, to our best knowledge this is the first rigorous COP model at the micro layer, which is self-contained and allows closed-form solutions. It opens the door to more sophisticated or realistic COP models in the future.

\item To the Algo practitioners on Wall Street, the model does reveal the intriguing dependency and competition among different key Algo components: aggressive sniping, passive waiting, execution cost, information leakage, market impact, and the requirement of completion. The modeling techniques here can always be tweaked to facilitate existing COP processes.
\end{itemize}

\subsection{The Placement Problem to be Modelled}
\label{subsec:2.3-problem}

Recall earlier in this section, after macro bucketing and micro binning, one ends up with a COP problem like the following:
\[
	\mbox{to buy 5 lots over a micro bin [1:00pm, 1:01pm],}
\]
or more generally, to buy $q$ lots over a micro bin $[T_0, T_1]$ with a brief duration $\Delta T = T_1 - T_0$  of 30 or 60 seconds or so. 

The COP problem modelled herein is formulated as follows. A given micro bin $[T_0, T_1]$ is partitioned into $N$ action times:
\[
	t_0 = T_0 < t_1 < \ldots < t_{N-1} < t_N = T_1. 
\]
In practice, one could choose periodic knots, say, $\tau=5$ or 10 seconds, and
\[
	t_{n+1} = t_n + \tau, \quad n=0, 1, \ldots, N-1.
\]
Actual implementation could also have them randomized for anti-gaming. Since the micro bin duration $\Delta T = T_1 - T_0 = t_N -t_0$ is brief, it is also assumed that the BBO, i.e., the best bid and offer, remain unchanged over the given micro bin. (This is introduced merely for convenience. In reality, the BBO can change realistically in the current model as long as one assumes that the benchmark market price is the moving mid price and that the limit order placement is cancelled and replaced whenever the BBO move so that it is effectively pegged to the BBO.)

Following the running example introduced earlier, we shall always work with a buy order - to buy $q_0$ lots over a micro bin $[T_0, T_1]$. For a {\em buy} order, we shall introduce the following concepts:
\begin{itemize}
\item the passive or near touch - the best bid of the venue, and

\item the aggressive or far touch - the best offer of the venue.
\end{itemize}
For a sell order the other way around holds. Furthermore, let $q_n$ denote the remaining lots right before $t \in [t_0, t_N]$. To simplify notation, we introduce the following convention: for any observable, variable or parameter $X$,
\[
	X_n := X_n^{-} =\lim_{\eps \to 0^+} X_{t - \eps}, \myand
	X_n^+ = \lim_{\eps \to 0^+} X_{t + \eps}. 
\]
For a continuous $X$ there is no difference among the three. For the proposed COP model, a sniping action (or a control in the context of dynamic programming) will take place right at a given action time $t_n$, and hence they could differ. 
\begin{mdframed}[backgroundcolor=orange!20] 
\noindent Attention - We have defaulted $X_n$ to $X_n^-$ to simplify notation and equation lines.
\end{mdframed}

The proposed COP model adopts the following action plan.
\begin{enumerate}[(i)]
\item At any time $t$, as long as $q_n >0$, a single-lot limit order will be placed at the passive touch. 

\item Whenever such a limit order is filled at $t$ and the remainder $q_t^+ = q_t -1$ is still positive, a new single-lot limit order will be immediately placed at the passive touch.

\item At each action time $t_n$ from
\[
	t_0, \ldots, t_n, \ldots, t_{N-1},
\]
if $q_n$ (which represents $q_n^-$) still has some positive lots to trade, the COP Algo has an option to send a single-lot market order at the far touch. We shall nickname this action by ``single-lot sniping" or simply ``sniping." The term ``Sniper" or ``Sniping" has been popularly used in the Algo world, e.g., the Sniper Algo of Credit Suisse in \href{https://www.reuters.com/article/businesspro-usa-algorithm-strategies-dc/snipers-sniffers-guerillas-the-algo-trading-war-idUSN3040797620070531}{this 2007 article of Reuters} (with an active URL link in PDF).
\end{enumerate}

The major three characteristics of this target COP problem are: order completion, passive waiting, and aggressive sniping, which are elaborated as follows. 

\paragraph{The Constraint on Completion}

Order completion is usually enforced at the macro layer. It is either explicitly formulated into optimization as a constraint (e.g., for the IS Algo) or enforced through hard scheduling (e.g., for TWAP/VWAP Algos). For micro-layer execution, e.g., executing 5 lots within a micro bin of 30 or 60 seconds, completion can be soft or delayed, in order to better dance with the liquidity waves in the market. The unfilled can be handed over to the next micro bin, and so on so forth. The macro Algo on the top usually deploys a dedicated schedule ``keeper" to enforce the schedules.

\paragraph{Passive Limit Order at the Near Touch}

Recall that for convenience, the BBO have been assumed invariant over a brief micro bin $[T_0, T_1]$. Let
\[
	HS = \frac 1 2 (BestOfferPrice - BestBidPrice)
\]
denote the half spread, and 
\[
	Mid = \frac 1 2 (BestOfferPrice + BestBidPrice)
\]
the unbiased mid price that represents the true market value (TMV) at the moment. Compared with other more sophisticated averaging schemes, this is usually called the simple mid. Once filled, a limit order saves a half spread $HS$ compared with the TMV. However, if left unfilled by the target end time, limit orders can jeopardize order completion. A reasonable COP model must be able to reflect this tradeoff.

\paragraph{Aggressive Sniping at the Far Touch}

Market or marketable orders take out liquidity from the top of the opposite LOB, i.e., the far touch. At the micro layer, market orders are always kept small, e.g., a couple of lots on average for major exchanges in US. Hence the proposed model always assumes that such small orders are filled instantaneously. Market orders fill fast and help achieve completion, but at the cost of a half spread $HS$. Furthermore, aggressive market orders could also leak information and result in opposite market participants or market makers biasing their perceived TMV towards the aggressive touch. As a result, they become less willing to take out ``our" limit orders posted at the passive touch. A reasonable COP model must be able to demonstrate such tradeoffs as well.

\section{The COP Model Based on Dynamic Programming}
\label{sec:3-model}

\subsection{Model and Process Assumptions}
\label{subsec:3.1-assumption}

We now introduce the basic assumptions for the model and process.

\paragraph{Assumptions on Limit Order Placement} 

For the placement of limit orders, the following basic assumptions are made.
\begin{enumerate}[(A)]
\item The size of the limit order is always a single lot.
\item A single-lot limit order is placed initially at $t_0$. 
\item Afterwards, whenever the limit order is taken out by an opposite market order at time $t$, a new single-lot limit order is immediately replenished as long as the remaining position $q_n^+ = q_n -1$ is still positive. 
\item For any time interval of duration $\Delta t$, as long as ``we" do not snipe at the aggressive touch within the interval, the number $W$ of single-lot limit orders being hit is subject to a Poisson distribution $N(\lambda \Delta t)$ with some rate $\lambda$. That is
\beginE \label{eqn:assumption-W-Poisson}
	\Prob(W =n) = e^{-\lambda \Delta t}   \frac{(\lambda\Delta t)^n}{n!}, \quad n=0, 1, \ldots.
\closeE
\end{enumerate}
It is also assumed that once being hit the entire lot is taken. 
Poisson distribution or process is not unfamiliar in the context of Algo trading and limit order books~\cite{algo_obizhaeva_wang13}. 

\paragraph{Assumptions on the Aggressive Market Orders}

For sniping using aggressive market orders, e.g., sending a marketable order $u_n$ at time $t_n$ at the far touch, the following assumptions are made.
\begin{enumerate}[(a)]
\item The ideal goal is to set $u_n$ to either a single lot or zero (i.e., no sniping), so that information leakage and market impact can be curbed. In reality, to facilitate a tractable dynamic programming (DP) formulation with closed-form solutions, the single lot constraint is not explicitly imposed. The cost function designed later will naturally encourage $u_n$ to stay small.

\item It is also assumed that once an aggressive market order $u_n$ is sniped, the entire order will be filled. Since the cost function in general keeps $u_n$ small, this assumption holds naturally for most liquid securities. 

\item Since during the short duration of a micro bin, the BBO are assumed to be invariant, we adopt the following model for information leakage caused by an aggressive sniping at time $t$. 

Once $u_n$ lots are taken out (by ``us") at the far touch, other market participants, including in particular market makers, will update their belief on the TMV and bias it towards the far touch. As a result,
\begin{mdframed}[backgroundcolor=orange!20] 
\noindent  either fewer opposite participants are willing to snipe at the passive touch or market makers will also cancel their more passive inside limit orders and replace with new ones at the passive touch.
\end{mdframed}
The latter will congest the queue at the passive touch. Hence heuristically both will reduce the chance of ``our" limit orders being hit by the market. 

Quantitatively, we assume that the Poisson hitting rate introduced in Eqn.~\eqref{eqn:assumption-W-Poisson} will be negatively impacted by the following linear form:
\beginE \label{eqn:assumption-Poisson-drop}
\lambda^+_n = \lambda_n - \eta   u_n,
\closeE 
after $u_n$ lots are sniped  and filled at time $t_n$.  Here $\eta > 0$ is a model parameter that calibrates the rate of information leakage or market impact caused by aggressive sniping. In general, it should depend on the liquidity profile of a given security. 

For instance, assume $\lambda = 5.0$ lots per minute, and $\eta = 0.5$ per lot per minute. Then a sniping of $u_n=2$ lots at some time $t_n$ will reduce the Poisson hitting rate according to:
\[
	\lambda^+_n = \lambda_n - \eta   u_n = 5.0 - 0.5 * 2 = 4.0.
\]
The impact parameter $\eta$ can be calibrated or estimated using experimental orders that are designed specifically for this purpose. 
\end{enumerate}

\subsection{State and Control Variables, and State Transition}
\label{subsec-3.2-state-transition}

To develop the dynamic programming (DP) COP model, we first define the state variables and their transition.

There are two state variables, $q_n$ and $\lambda_n$, or organized into a state vector $\bx_n=(q_n, \lambda_n)^T$, where the superscript $T$ denotes transposition of vectors or matrices.
\begin{itemize}
\item State variable $q_n$ denotes the outstanding lots still needed to be traded right before $t_n$.  The symbol is equivalent to $q_n^-$ but with the minus superscript omitted, as set up earlier.

\item State variable $\lambda_n$ denotes the Poisson hitting rate right before $t_n$. It represents the rate the opposite aggressive orders hit ``our" limit orders. It changes whenever ``we" snipe a market order $u_n$ at the far touch, due to information leakage and its digestion by other market participants.
\end{itemize}

The control variable is the aggressive takeout order $u_n$ that ``we" snipe at $t_n$ at the far touch. Furthermore, let $W_n$ denote the Poisson random number of ``our" limit orders being hit by the opposite aggressive orders. Then from $t_n$ to $t_{n+1}$ (or more precisely $t_{n+1}^-$), the following state transition equations hold.
\begin{align}
q_{n+1}  		&= q_n - u_n - W_n, 		\label{eqn:state-transition-q}\\
\lambda_{n+1}	&= \lambda_n - \eta u_n. 	\label{eqn:state-transition-lam}
\end{align}
In the vector form, define
\[
	\ba = (1, \eta)^T, \qquad \bz_n = (W_n, 0)^T. 
\]
Then the state vector transits as follows:
\beginE \label{eqn:state-transi-vector}
	\bx_{n+1} = \bx_n - \ba   u_n - \bz_n.  
\closeE

\subsection{The Stage Cost}
\label{subsec:3.2-state-cost}

All variables are assumed to be continuous, as normally done in the Algo literature. (In reality trades are mostly in whole shares or lots.) 

Following all the previous preparation, the COP problem is formulated as a DP problem with controls taken at one of the following action times:
\[
	t_n \in \{ t_0, t_1, \ldots, t_{N-1} \}.
\]
No action is taken at the terminal time knot $t_N$. At each $t_n$, we first define the initial candidate $j_n^{(1)}$ for the stage cost, which is still subject to revision later on:
\beginE \label{eqn:stage-cost-raw-A}
	j_n^{(1)}( u_n, W_n \mid \bx_n ) = \gamma_n   q_n^2 + u_n - W_n.
\closeE
It is explained as follows.
\begin{enumerate}[(a)]

\item The first term $\gamma_n q_n^2$ favors fast execution so that $q_n$'s quickly touch down to zero. It appears in earlier works for both static and dynamic macro Algos, e.g., Algren-Chriss~\cite{algo_almgren_chriss00}, Hora~\cite{algo_hora2006_lqr}, and Shen~\cite{algo_shen15_pretrade,algo_shen17_lqrIS}, just to name a few. The penalty coefficient $\gamma_n$'s are the control parameters to penalize trading delays. In general, $\gamma_n$ can be set in proportion to the real-time variance $\sigma_n^2$ of the secuity, i.e., in the form of $\gamma_n = \tilde{\gamma}_n    \sigma_n^2$. In addition, $\gamma_n$ should increase monotonically to facilitate order completion. For instance, $\gamma_N=+\infty$ would enforce hard completion: $q_N=0$.

\item $W_n$ is the number of single-lot limit orders being taken out by opposite market orders at ``our" passive touch. Limit orders save a half-spread $HS$. We incorporate the scaling of $HS$ into $\gamma_n$ so that $-HS \cdot  W_n$ can be simplified to $-W_n$. Following the general Poisson setting in Eqn.~\eqref{eqn:assumption-W-Poisson}, we assume more specifically that $W_n$ is subject to the Poisson distribution $N(\lambda_n^+ \Delta t_n)$ with
\[
	\lambda_n^+ = \lambda_n - \eta u_n, \myand \Delta t_n = t_{n+1} - t_n. 
\]

\item $u_n$ is the number of lots that ``we" snipe at the aggressive touch at the action time $t_n$. It pays the cost of a half spread, i.e., $HS \cdot u_n$. Since $HS$ is scaled into $\gamma_n$, it is simply expressed as $u_n$ in the stage cost. 

For a buy order, $u_n$ is preferably nonnegative. In order to facilitate a close-form solution, however, we do not explicitly impose the constraint of $u_n \ge 0$. When $u_n < 0$, we shall interpret it as an aggressive sell order of size $-u_n$ at the current passive touch. If this is the case, the update position will increase:
\[
	q_n^+ = q_n - u_n > q_n. 
\]
The first term $\gamma_n q_n^2$ will discourage such opposite trades as long as the risk aversion weights $\gamma_n$'s are not negligible. 

On the other hand, compared with the market mid price, selling at the passive touch also incurs a cost of a half spread, i.e, $HS \cdot (-u_n)$. Hence when $u_n$ is allowed to be signed, the stage cost in Eqn.~\eqref{eqn:stage-cost-raw-A} should at least be revised to:
\[
	j_n^{(2)}( u_n, W_n \mid \bx_n ) = \gamma_n   q_n^2 + |u_n| - W_n.
\]
Since we intend to design a DP model with a closed-form solution, the absolute value is further revised to a squared form:
\beginE \label{eqn:stage-cost-final}
	j_n( u_n, W_n \mid \bx_n ) = \gamma_n   q_n^2 + u_n^2 - W_n,
\closeE
which is the final stage cost adopted for the current model. 

When $u_n$ stays close to a single lot, $u_n^2 \simeq |u_n|$. For $|u_n|> 1$, this quadratic form penalizes big sizes even heavier than the linear form. It favors smaller aggressive order sizes as a result.

Inspired by the $\Gamma$-convergence theory and its application in multi-phase variational problems~\cite{bImage_ChanShen}, one could also introduce the double-well cost function:
\[
		\frac{u_n^2 (1 - u_n)^2}{\eps}, \mywith 0 < \eps \ll 1.
\]
This will softly enforce the binary sniping behavior - either no action with $u_n=0$ or sniping with a single lot $u_n =1$. However, such high order non-convex costs can completely thwart the effort of designing a DP model with a unique and close-form solution. 

\end{enumerate}

\begin{mdframed}[backgroundcolor=orange!20] 
\noindent \textbf{Summary.\;} The stage cost model in Eqn.~\eqref{eqn:stage-cost-final} and the state transition model in Eqn.~\eqref{eqn:state-transi-vector} define the proposed stochastic dynamic programming model for child order placement (COP).
\end{mdframed}


\section{Bellman Equations and Optimal Solutions}
\label{sec:4-solution}

\subsection{Value Functions $V(\bx)$}

At any ``current" action time $t_n$ with state variable $\bx_n = (q_n, \lambda_n)^T$, for any choice of policy or action sequence $\bu_n = (u_n, \ldots, u_{N-1})$ at $(t_n, \ldots, t_{N-1})$, let $\bW_n = (W_n, \ldots, W_{N-1})$ denote the resulting Poisson hits on ``our" passive limit orders during individual intervals $[t_k, t_{k+1})$'s.  Then the future cost is given by
\beginE
\begin{split}
  J_n( \bu_n, \bW_n \mid \bx_n) 
  		& = J_n(u_n, \ldots, u_{N-1}; W_n, \ldots, W_{N-1} \mid \bx_n ) 	\\
 		& = j_n(u_n, W_n \mid \bx_n) + \ldots + j_{N-1}(u_{N-1}, W_{N-1} \mid \bx_{N-1}) + j_N(\bx_N) \\
 		& = j_n(u_n, W_n \mid \bx_n) + J_{n+1}(\bu_{n+1}, \bW_{n+1} \mid \bx_{n+1}) \\
\end{split}
\closeE
Here $j_N(\bx_N)$ denotes the terminal cost at $t_N$. 
Define the value function by:
\beginE \label{eqn:value-V}
	V_n(\bx_n) = \inf_{\bu_n} \EXP{\bW_n}{J_n(\bu_n, \bW_n \mid \bx_n)},
\closeE
ranging over all state-driven policies in the form of $u_k = \phi_k(\bx_k)$, $k=n, \ldots, N-1$. 
Then we have the Bellman equation at each action time $t_n$:
\beginE \label{eqn:Bellman}
	V_n(\bx_n) = \inf_{u_n} \EXP{W_n}{j_n(u_n, W_n \mid \bx_n)}
					+ \EXP{W_n}{V_{n+1}(\bx_{n+1})}.
\closeE

The terminal cost at $t_N$ is defined to be
\beginE \label{eqn:terminal-V}
	V_N(\bx_N) = j_N(\bx_N) = \gamma_N   q_N^2. 
\closeE
The other two terms are dropped out since the end of the given micro bin is reached. 

Next, we shall work out first the solution for the last period $[t_{N-1}, t_N)$ to gain some tangible knowledge, and then the general solution in the framework of stochastic LQR.

\subsection{Solution at the Last Period}

At action time $t_{N-1}$ for the last period $[t_{N-1}, t_N)$, the Bellman equation reads:
\[
	V_{N-1}(\bx_{N-1}) 
		= \inf_{u_{N-1}} \; \gamma_{N-1} q_{N-1}^2 + u_{N-1}^2 - \lambda_{N-1}^+ \Delta t_{N-1}
						+ \gamma_N \EXP{W_{N-1}}{(q_{N-1} - u_{N-1} - W_{N-1})^2}.
\]	
For clarity in calculation, we drop the subscript $N-1$ so that it becomes
\[
	V(\bx) 	= \inf_{u} f(u \mid \bx) 
			= \inf_u \; \gamma q^2 +u^2 - \lambda^+ \dt + \gamma_N \E_W (q - u - W)^2,
\]
with the current state vector $\bx =(q, \lambda)^T$ which is known, and $\lambda^+ =  \lambda - \eta u$. The problem becomes the minimization of a single-variate function $f(u \mid \bx)$ given $\bx$. 

Since $\EXP{}{W}=\lambda^+\dt$, and $\E[X^2] = \Var(X) + \E[X]^2$ for a generic random variable $X$, one has
\[
\begin{split}
	\E(q - u - W)^2 & = \Var( q-u-W ) + (q - u - \lambda^+\dt)^2 \\
					& = \Var(W) + (q - u - \lambda^+\dt)^2 \\
					& = \lambda^+\dt + (q - u - \lambda^+\dt)^2. \\
\end{split}
\]
As a result, the derivative of $f$  is:
\[
\begin{split}
	\frac{df}{du} &= 2u + (\gamma_N -1) \dt \frac{d\lambda^+}{du}
				 - 2 \gamma_N (q - u - \lambda^+\dt)(1 +\dt \frac{d\lambda^+}{du})  \\
				  &= 2(1 + \gamma_N(1-\eta\dt)^2) u - \eta\dt(\gamma_N-1) - 2 \gamma_N(1-\eta\dt)(q-\lambda\dt). \\
\end{split}
\]
At the optimal $u_\ast$, the derivative vanishes. Hence the optimal aggressive trading is given by:
\beginE 
\label{eqn:ustar-last-period}
\begin{split}
	u_\ast 		&= \alpha_\ast + \beta_\ast (q - \lambda\dt), \mywith \\
	\alpha_\ast &= \frac{(\gamma_N-1)\eta\dt}{2(1+\gamma_N(1-\eta\dt)^2)}, \myand
	\beta_\ast 	= \frac{\gamma_N(1-\eta\dt)}{1 + \gamma_N(1-\eta\dt)^2}
\end{split}
\closeE
In particular, the optimal policy $u_\ast = \phi_\ast(\bx) = \phi_\ast(q, \lambda)$ is a linear function of the state variable. 

Consider the asymptotic case when $\gamma_N = + \infty$. Then one has
\[
	\alpha_\ast = \frac{\eta\dt}{2(1-\eta\dt)^2}, \myand 
	\beta_\ast 	= \frac{1}{1-\eta\dt}.
\]
In particular, for a highly liquid security so that $\eta \simeq 0$, the optimal policy is simply
\[
	u_\ast = \alpha_\ast + \beta_\ast (q - \lambda\dt) \simeq q - \lambda\dt.
\]
That is, on the last action time $t_{N-1}$, the proposed COP Algo will trade the expected remaining lots that cannot be filled by the passive limit orders $W$ (since $\E[W] = \lambda^+\dt \simeq \lambda\dt$ when $\eta \simeq 0$). This certainly makes business sense. 

In terms of the value function at $t_{N-1}$, one then has
\beginE 
\label{eqn:value-last-step}
\begin{split}
V(\bx)  &= f( u_\ast \mid \bx ) \\
		&= \gamma q^2 + u_\ast^2 + (\gamma_N-1)(\lambda\dt - \eta\dt u_\ast) 
				+ \gamma_N ( q - \lambda\dt - (1-\eta\dt)u_\ast )^2 \\
		&= \gamma q^2 + c_1(q-\lambda\dt)^2 + c_2 (q-\lambda\dt) + c_3,
\end{split}
\closeE
where 
\[
	c_1 = \beta_\ast^2 + \gamma_N ( 1 - (1 - \eta\dt)\beta_\ast)^2 
		= \frac{\gamma_N}{1 + \gamma_N (1 - \eta\dt)^2} > 0.
\]
Hence the value function for the last period can be written in the canonical quadratic form as:
\beginE 
\label{eqn:value-last-vector-form}
	V(\bx) = \bx^T  P   \bx + \bb^T   \bx + c,
\closeE
where $P$ must be a positive definite matrix. This is because  by the quadratic portion of $V(\bx)$,
\[
	\gamma q^2 + c_1(q-\lambda\dt)^2  = 0 \Rightarrow q = 0, \; \lambda = 0.
\]

Next we show that this is not accidental.


\subsection{General Solution to the Stochastic LQR}
\label{subsec:general-solution}

In general, for $n=0, 1, \ldots, N-2$, assume that the value function at $n+1$ is in the quadratic form:
\beginE 
\label{eqn:value-n-next}
V_{n+1}(\bx_{n+1}) = \bx_{n+1}^T   P_{n+1}   \bx_{n+1} 
						+ \bb_{n+1}^T   \bx_{n+1} + c_{n+1},
\closeE
where $P_{n+1}$ is   positive definite. We now show that this implies that
\[
	V_n(\bx_n) = \bx_n^T   P_n   \bx_n + \bb_n^T   \bx_n + c_n,
\]
where $P_n$ is also   positive definite. 

Furthermore, we show that the optimal action is given in the linear form:
\[
	u_n = \phi_n(\bx_n) = \alpha_n + \bbeta_n^T   \bx_n. 
\]
The objective is to derive $\alpha_n, \bbeta_n, P_n, \bb_n$ and $c_n$ from $P_{n+1}, \bb_{n+1}$ and $c_{n+1}$ recursively. 

For clarity, we drop the subscript $n$ so that for any variable or parameter $X$, we use instead
\[
	X_n \longrightarrow X, \myand X_{n+1} \longrightarrow X_1. 
\]
In particular, the value function at $n+1$ now assumes the cleaner form:
\beginE
\label{eqn:value-at-1}
V_1(\bx_1) = \bx_1^T   P_1   \bx_1 + \bb_1^T   \bx_1 + c_1.
\closeE
Also the state transition equation becomes:
\[
	\bx_1 = \bx - \ba u - \bz, \mywith \ba=(1, \eta)^T, \quad \bz = (W, 0)^T,
\]
where $\eta$ is a model parameter or constant that represents the market impact or information leakage and $W$ is the Poisson random hits on ``our" passive limit orders over $[t_n, t_{n+1})$.

Then the value function at $t_n$ is given by:
\[
	V(\bx) = \min_u f(u \mid \bx) = \min_u \gamma q^2 + u^2 - \lambda^+ \dt 
				+ \E_W{V_1(\bx_1)}.
\]	
Let $p_{11}$ denote the (1,1)-element of $P_1$, and $\bz_\E =(\lambda^+\dt, 0)^T = \E[\bz]$. Then 
\[
\begin{split}
	\E_W V_1(\bx_1) &= V_1(\E_W \bx_1) + \E_W (\bz-\bz_\E)^T   P_1   (\bz-\bz_\E) \\
					&= V_1(\bx - \ba u - \bz_\E) + p_{11} \Var(W) \\
					&= V_1(\bx - \ba u - \bz_\E) + p_{11} \lambda^+\dt.
\end{split}
\]
We now Define
\beginE 
\label{eqn:assisting-matrices}
	L = \begin{pmatrix}
	0 & \dt \\
	0 & 0
	\end{pmatrix}, \quad
	J = I_2 - L, \myand
	\bh = \begin{pmatrix}
			1-\eta\dt \\
			\eta
		  \end{pmatrix},
\closeE
where $I_2$ denote the 2 by 2 identity matrix. Then 
\[
\begin{split}
\bz_\E &= \begin{pmatrix}
		\lambda\dt - \eta\dt u \\ 0
		\end{pmatrix} = L \bx - \begin{pmatrix} \eta\dt \\ 0 \end{pmatrix} u, \\
\bx - \ba u - \bz_\E &= J   \bx - \bh u.
\end{split}
\]
Hence we have, with $l_{11}=1-p_{11}$, 
\beginE
\label{eqn:f(u)}
	f(u) = f(u \mid \bx) = \gamma q^2 + u^2 - l_{11} \lambda^+\dt  + V_1(J  \bx - \bh u). 
\closeE
Assume that
\beginE \label{eqn:f-df-ABC}
	f(u) = f(0) - Bu + Au^2. \qquad \mathrm{Then,} \quad f'(u) = 2A u - B.
\closeE
On the other hand, direct differentiation gives
\[
\begin{split}
f'(u) 	&= 2u + l_{11} \eta\dt - \bh^T   \grad V_1(J  \bx - \bh u) \\
		&= 2(1 + \bh^T   P_1   \bh) u - (\bh^T   \bb_1 - l_{11}\eta\dt)
			 - 2 \bh^T   P_1   J   \bx.
\end{split}
\]
Hence we have
\beginE \label{eqn:coeff-A-B}
	A = 1 + \bh^T   P_1   \bh, \myand 
	B = (\bh^T  \bb_1  - l_{11}\eta\dt) + 2\bh^T   P_1   J   \bx. 
\closeE
Therefore, the optimal policy at $t_{n-1}$ is given by
\beginE
\label{eqn:optimal-policy-n}
\begin{split}
 	u_\ast 		&= \frac{B}{2A} = \alpha_\ast + \bbeta_\ast^T \bx, \mywith \\
 	\alpha_\ast &= \frac{\bh^T  \bb_1  - l_{11}\eta\dt}{2(1 + \bh^T   P_1   \bh)}, \myand
 	\beta_\ast^T 	= \frac{\bh^T   P_1   J }{ 1 + \bh^T   P_1   \bh}
\end{split}
\closeE

Next we derive the associated value function $V(\bx)=f(u_\ast \mid \bx)$. For convenience, for any positive definite matrix $Q$ and any real vector $\bx$ of the same dimension, we define the Q-stretched Euclidean norm by:
\[
	\| \bx \|_Q^2 = \bx^T   Q   \bx. 
\]
Then the coefficient $A$ is simply $1 + \|\bh\|_{P_1}^2$.

Since $2A u_\ast = B$, one has $2 Au_\ast^2 = Bu_\ast$.  Hence by Eqn.~\eqref{eqn:f(u)},
\[
\begin{split}
 V(\bx)     &=  f(u_\ast \mid \bx) 	\\
 			&= f(0) - B u_\ast + A u^2_\ast \\
 			&= f(0) - A u_\ast^2 \\
 			&= \gamma q^2 - l_{11}\lambda\dt +V_1(J  \bx) -A (\alpha_\ast + \bbeta_\ast^T   \bx)^2 \\
 			&= \gamma q^2 - l_{11}\lambda\dt + (\bx^T J^T P_1 J \bx + \bb_1^T J \bx + c_1)
 				- ( 1 + \|\bh\|_{P_1}^2 )(\alpha_\ast + \bbeta_\ast^T \bx)^2 \\
 			&= \gamma q^2 + \bx^T (J^T P_1 J) \bx 
 					- (1 + \|\bh\|_{P_1}^2 ) \bx^T \bbeta_\ast \bbeta_\ast^T \bx \\
 			&\qquad - l_{11}\lambda\dt + \bb_1^T J \bx 
 					- 2 (1 + \|\bh\|_{P_1}^2) \alpha_\ast\bbeta_\ast^T \bx \\
 			&\qquad + c_1 - (1 + \|\bh\|_{P_1}^2) \alpha_\ast^2  \\
 			&= \bx^T P \bx + \bb^T \bx + c,
\end{split}
\]
where the quadratic parameters are given by
\beginE 
\label{eqn:value-P-b-c}
\begin{split}
 P 		&= \gamma \beginPM 1 & 0 \\ 0 & 0 \closePM + J^T Q J, \mywith
 			Q = P_1 - \frac{P_1 \bh \bh^T P_1}{1 + \|\bh\|_{P_1}^2},	 \\
 \bb^T 	&= \bb_1^T J - l_{11}\dt(0, 1) - 2 ( 1 + \|\bh\|_{P_1}^2) \alpha_\ast \bbeta_\ast^T,	 \\
 c		&= c_1 - ( 1 + \|\bh\|_{P_1}^2 ) \alpha_\ast^2. 
\end{split}
\closeE
where $l_{11}=1-P_1(1,1)$ and $\alpha_\ast, \bbeta_\ast$ are given as in Eqn.~\eqref{eqn:optimal-policy-n}.

We now show that $P$ is   positive definite.
\begin{lemma}
If $P_1$ is  positive definite, so must be $P$. 
\end{lemma}

\beginProof
Since $J = I_2 - L$ is non-singular as in Eqn~\eqref{eqn:assisting-matrices}, it suffices to show that $Q$ is positive definite. 

For any non-zero vector $\bv\in \realR^2$, previously we have used the notation $\|\bv\|_{P_1}$ to denote the $P_1$-stretched Euclidean distance. More generally, we use
\[
	\inner{\bv}{\bh}_{P_1} : = \bv^T P_1 \bh
\]
to denote the $P_1$-stretched inner product. By the Cauchy-Schwarz Theorem, one has
\[
	\inner{\bv}{\bh}_{P_1} \le \|\bv\|_{P_1} \cdot \|\bh\|_{P_1}. 
\]
Then back to $Q$, one has
\[
\begin{split}
 \bv^T Q \bv &= \bv^T \left( P_1 - \frac{P_1 \bh \bh^T P_1}{1 + \|\bh\|_{P_1}^2} \right) \bv \\
 			 &= \|\bv\|_{P_1}^2 - \frac{\inner{\bv}{\bh}_{P_1}^2}{1 + \|\bh\|_{P_1}^2} \\
 			 &= \frac{\|\bv\|_{P_1}^2 + \|\bv\|_{P_1}^2 \|\bh\|_{P_1}^2 - \inner{\bv}{\bh}_{P_1}^2}{1 + \|\bh\|_{P_1}^2} \\
 			 &\ge \frac{\|\bv\|_{P_1}^2}{1 + \|\bh\|_{P_1}^2} > 0.
\end{split}
\]
Since this holds for any non-zero vector $\bv$, $Q$ and hence $P$ must be positive definite. 
\closeProof

We have thus established the following theorem, with $J_n$ and $\bh_n$ defined as in Eqn.~\eqref{eqn:assisting-matrices}:
\[
	J_n = \beginPM
	1 & -\dt_n \\
	0 & 1
	\closePM, \myand
	\bh_n = \begin{pmatrix}
			1-\eta\dt_n \\
			\eta
		  \end{pmatrix}.
\]
They are constant for equal partitioning when $\dt_n=t_{n+1} - t_n$'s are all the same. 

\begin{mdframed}[backgroundcolor=orange!8]
\begin{theorem} \label{thm:solution-COP}
Let $V_N(\bx_N) = \gamma_N q_N^2$ be the terminal cost at the ending time $t_N$. Then at each action time $t_n$ with $n<N$, there exist a positive definite 2 by 2 matrix $P_n$, a 2 by 1 vector $\bb_n$, a scalar $c_n$, such that the value function $V_n$ is given by:
\beginE
\label{eqn:value_function_Vt}
 	V_n(\bx_n) = \bx_n^T P_n   \bx + \bb_n^T  \bx_n + c_n 
 				 = (q_n, \lambda_n)   P_n   \beginPM q_n \\ \lambda_n \closePM 
 				 	+ \bb_n^T \beginPM q_n \\ \lambda_n \closePM + c_n.
\closeE
The optimal policy $u_n$ is given by the linear form using parameters at $t_{n+1}$ :
\beginE
\label{eqn:optimal_u_n}
\begin{split}
	u_n^\ast 		&= \phi_n(\bx_n) =  \alpha_n^\ast + \bx_n^T \bbeta_n^\ast, \mywith \\
	\alpha_n^\ast 	&= \frac{\bb_{n+1}^T \bh_n  - (1-P_{n+1}(1,1))\eta\dt_n}{2(1 + \bh_n^T   P_{n+1}   \bh_n)}, \myand  
	\bbeta_n^\ast = \frac{J_n^T   P_{n+1}   \bh_n }{ 1 + \bh_n^T   P_{n+1}   \bh_n},
\end{split}
\closeE
where $P_{n+1}(1,1)$ denotes the (1,1)-element of $P_{n+1}$. 
Furthermore, the structure of the value functions also cascades backwards as follows:
\beginE
\label{eqn:value-back-cascading}
\begin{split}
 P_n 		&= \gamma_n \beginPM 1 & 0 \\ 0 & 0 \closePM 
 				+ J_n^T \left( P_{n+1} - \frac{P_{n+1} \bh_n \bh_n^T P_{n+1}}{1 + \bh_n^T P_{n+1}\bh_n}\right) J_n,  \\
 \bb_n 		&=  J_n^T \bb_{n+1} - (1-P_{n+1}(1,1))\dt_n \beginPM 0 \\ 1 \closePM
 			 - 2 (1 + \bh_n^T P_{n+1}\bh_n) \alpha_n^\ast \bbeta_n^\ast,	 \\
 c_n		&= c_{n+1} - ( 1 + \bh_n^T P_{n+1}\bh_n ) (\alpha_n^\ast)^2, 
\end{split}
\closeE
with $n=N-1, \ldots, 1, 0$, and terminal values $P_N=\mathrm{diag}(\gamma_N, 0), \bb_N=\mathbf{0}$ and $c_N=0$.
\end{theorem}
\end{mdframed}

\newpage
\section{Conclusion and Disclaimers}
\label{sec:disclaimers}

We conclude the current work with the following comments and disclaimers. 

\begin{enumerate}[(1)]
	\item The proposed dynamic programming COP Algo had not been the internal or external product of any execution houses where the author worked previously. Any potential industrial conflict or suspected proprietary trespass should be promptly directed to the attention of the author, together with necessary evidences.

	\item In the spirit of reductionism and the pursuit of a dynamic programming COP model with closed-form solutions, the current model does not address other important execution or implementation details, including fragmented venues in the national market system (NMS), different trading sessions and rules, various order types, lit vs. dark, and so on. 
	
	\item The current work focuses exclusively on the dynamic interplay between aggressive takeout  orders and passive limit orders. The price improvement or cost is represented by a half spread. Information leakage or the market impact of aggressive orders is reflected in the reduction of Poisson hitting rates on passive limit orders.
	
	\item Like some earlier DP macro Algos, risk aversion is implemented by the delay cost in the stage cost model. It facilitates soft completion of a given child order over its designated micro time bin. Hard completion or catchup is usually implemented at the macro layer. 
	
	\item If the results here are to be integrated into an existing COP program in an execution house, a practitioner should apply some heuristic but necessary overlays. For instance, $u^\ast_n \leq 0$ can be interpreted as no sniping, while $u^\ast_n>0$ should also be capped by the  outstanding position $q_n$.  
	
	\item Overall, the author wishes that the current model could inspire more similar and rigorous works that can improve the heuristic decision trees prevailing in the COP processes in the contemporary Algo industry. 

\end{enumerate}

\section*{Acknowledgments}

Jackie Shen is very grateful to all the colleagues at the electronic or algorithmic trading desks of J.P. Morgan, Barclays and Goldman Sachs, esp. to many of our hard-working IT, RISK, and COMPLIANCE colleagues whose names hardly appear in the headlines, for their daily professional assistance as well as generous personal support.  Reliable and healthy electronic trading would be impossible without solid IT implementations of databases, data streaming, servers, networks, and multiple inter-dependent Algo components, or effective risk management and compliance controls.

This work was completed when the pandemic Covid-19 was sweeping through the entire globe mercilessly. Under tremendous mental pressure living in the epicenter of New York, the author is extremely grateful to his family, friends, and colleagues, as well as thousands of courageous and selfless medical professionals, policemen and policewomen, and fire fighters of this great city. 

The pandemic has actually brought the people in the city and around the globe much closer and more united, as my 9-year old observes from her numerous Zoom online classes and chats, as well as all the touching stories around the world on fighting against the virus. Beyond A.I. or automated trading ``robots" as the current work has covered, the pandemic has grounded all of us to the very core meaning of human beings and human societies. At the end of this darkest storm there will be a brightest rainbow --- so colorful, refreshing, and full of new hopes.

\end{document}